\documentclass[twocolumn,prl,preprintnumbers,superscriptaddress,amsmath,amssymb,english]{revtex4}
\usepackage{amsmath,amssymb,amsfonts,mathrsfs}
\usepackage{amsthm}
\usepackage{CJK}
\usepackage{bm}
\usepackage{babel}
\usepackage{graphicx}
\allowdisplaybreaks
\usepackage{braket}
\usepackage[breaklinks]{hyperref}
\usepackage{color}
\hypersetup{colorlinks=true, linkcolor=blue, citecolor=blue, filecolor=blue, urlcolor=blue}
\UseRawInputEncoding
\begin{document}

\title{Photoinduced Quantum Anomalous Hall States in the Topological Anderson Insulator}

\author{Zhen Ning}
\affiliation{Institute for Structure and Function $\&$ Department of Physics, Chongqing University, Chongqing 400044, People¡¯s Republic of China}
\affiliation{Chongqing Key Laboratory for Strongly Coupled Physics, Chongqing 400044, People¡¯s Republic of China}

\author{Baobing Zheng}
\affiliation{College of Physics and Optoelectronic Technology $\&$ Advanced Titanium Alloys and Functional Coatings Cooperative Innovation Center, Baoji University of Arts and Sciences, Baoji 721016, People¡¯s Republic of China}
\affiliation{Institute for Structure and Function $\&$ Department of Physics, Chongqing University, Chongqing 400044, People¡¯s Republic of China}
\affiliation{Chongqing Key Laboratory for Strongly Coupled Physics, Chongqing 400044, People¡¯s Republic of China}

\author{Dong-Hui Xu}
\email{donghuixu@hubu.edu.cn}
\affiliation{Department of Physics, Hubei University, Wuhan 430062, People¡¯s Republic of China}

\author{Rui Wang}
\email{rcwang@cqu.edu.cn}
\affiliation{Institute for Structure and Function $\&$ Department of Physics, Chongqing University, Chongqing 400044, People¡¯s Republic of China}
\affiliation{Chongqing Key Laboratory for Strongly Coupled Physics, Chongqing 400044, People¡¯s Republic of China}
\affiliation{Center for Quantum Materials and Devices, Chongqing University, Chongqing 400044, People¡¯s Republic of China}

\date{\today}

\begin{abstract}
	
The realization of the quantum anomalous Hall (QAH) effect without magnetic doping attracts intensive interest since magnetically doped topological insulators usually possess inhomogeneity of ferromagnetic order. Here, we propose a different strategy to realize intriguing QAH states arising from the interplay of light and non-magnetic disorder in two-dimensional topologically trivial systems. By combining the Born approximation and Floquet theory, we show that a time-reversal invariant disorder-induced topological insulator, known as the topological Anderson insulator (TAI), would evolve into a time-reversal broken TAI and then into a QAH insulator by shining circularly polarized light. We utilize spin and charge Hall conductivities, which can be measured in experiments directly, to distinguish these three different topological phases. This work not only offers an exciting opportunity to realize the high-temperature QAH effect without magnetic orders, but also is important for applications of topological states to spintronics.

\end{abstract}

\maketitle
The investigation of the topological phase of matter has been a major subject in condensed matter physics
\cite{Haldane,TKNN,Hasan,QiZhang}. Among various topological phases, the quantum anomalous Hall (QAH) insulator with the spontaneous time-reversal  ($\mathcal{T}$) symmetry breaking  in the absence of external magnetic fields \cite{Haldane} is especially interesting.  Dissipationless chiral edge transport with a quantized Hall conductance in the QAH insulator provides great potential in applications of the next generation of spintronic devices
with low dissipation. Another important example is the quantum spin Hall (QSH) insulator \cite{KaneMele1,KaneMele2,SCZhang}, also known as a two-dimensional topological insulator (TI). The QSH insulator is characterized by the $\mathcal{T}$ symmetry protected helical edge states and a $\mathbb{Z}_2$ index, which can be viewed as two copies of QAH insulators with opposite spin polarization. The experimental evidence of the QSH state was first confirmed in HgTe quantum wells with an inverted band structure \cite{TIexp}. Shortly after the discovery of the QSH state, inverted HgTe quantum wells were predicted to give rise to QAH states through the introduction of doping magnetic Mn atoms \cite{mdopping1}, but remain elusive experimentally. Alternatively, the QAH effect has been observed experimentally in magnetically doped topological insulator thin films \cite{Chang}. However, the observation of QAH effect requires ultralow temperatures due to inhomogeneity of ferromagnetic order. Therefore, for purposes of practical application, it is highly desirable for realizing high-temperature QAH effects.

As is well known, in QSH insulators, helical edge states can be robust against non-magnetic disorder, which also plays an important role in observing the helical edge transport in a realistic topological system \cite{DRRsidoped}. More strikingly, such disorder can endow a system with non-trivial topological states, i.e., the topological Anderson insulator (TAI) \cite{TAI}, in which disorder gives rise to the negative mass of Dirac quasi-particles and thereby drives the band inversion from the point of view of the mean field theory~\cite{Groth}. Recently, the TAI phase was experimentally observed in disordered atomic wires~\cite{expTAI} and photonic crystal systems \cite{PTAI}. Meanwhile, a light field provides a controllable approach to artificially manipulating topological states and their phase transition ~\cite{Oka1,Inoue,Lindner1,McIver, Liu2018Phosphorus,Deng2020carbon,Wang2018FeSe}.
Moreover, the marriage of two ingredients, disorder and light, has been used to produce and manipulate various topological phases recently \cite{Titum1,Titum2,Titum3,Roy,Yap,Stability,LDu,RChen2,Lindner3}. While recent advancements have been very encouraging, most studies of topological states in the presence of disorder and light fields are limited to spin-nonpolarized cases, thus making that spin-polarized topological phases and especially the QAH phase in systems induced by disorder and light have rarely been reported. To avoid mishaps of long-range ferromagnetic order suppressed by magnetic doping, one would like to search for a QAH mechanism in the non-magnetic TAI.


\begin{figure}[htb]
    \centering
    \includegraphics[width=0.46\textwidth]{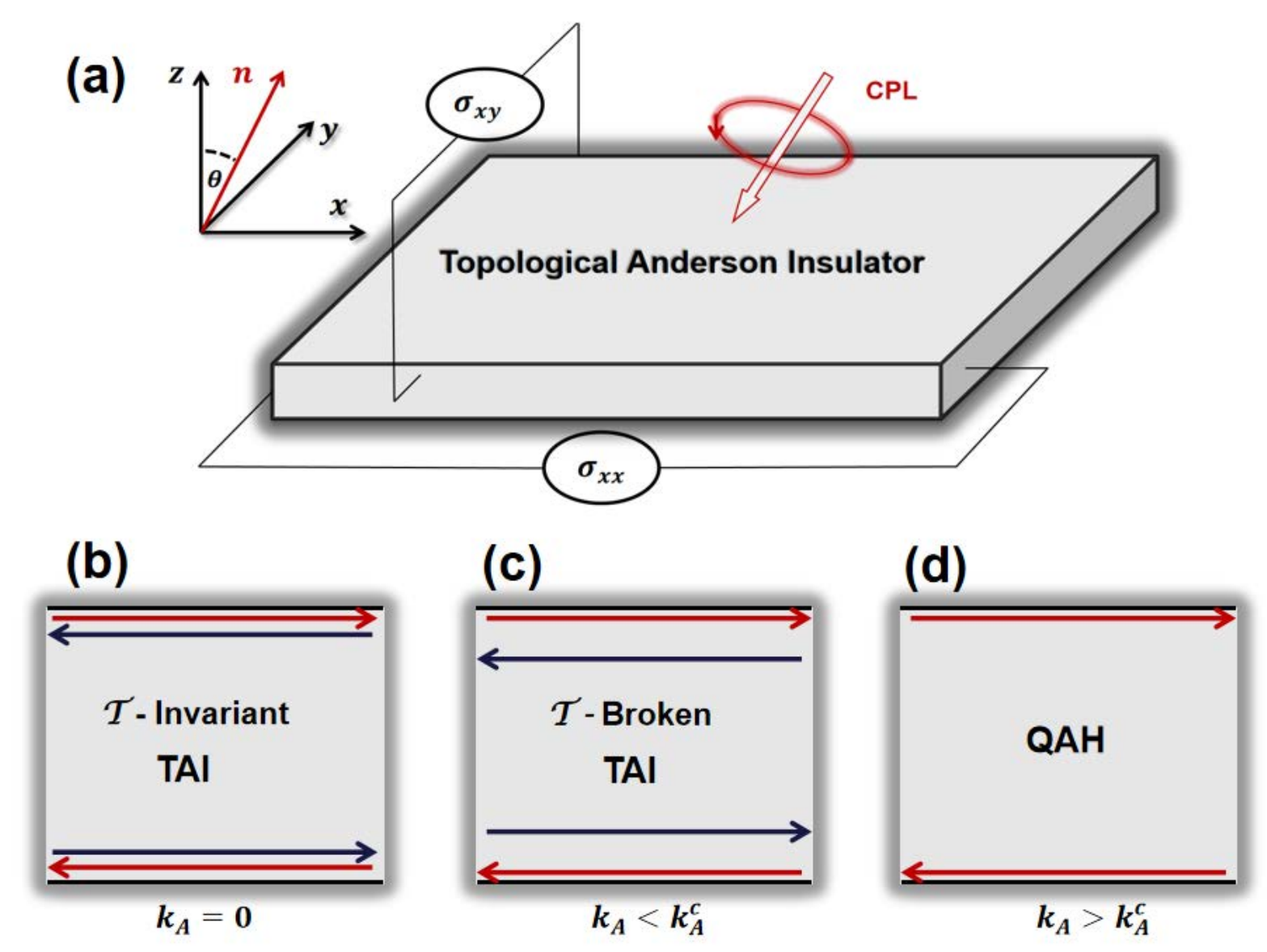}%
    \caption{
    (a) The schematic diagram of the TAI under irradiation of CPL.
    The direction of incident light is described as $\boldsymbol{n}=(\sin\theta\cos\varphi,\sin\theta\sin\varphi,\cos\theta)$ with $\theta$ (polar angle) and $\varphi$ (azimuthal angle) in the spherical coordinate. (b) Without irradiation of CPL (i.e., the light intensity $k_A = 0$), the TAI support a pair of counterpropagating edge states with the Kramers degeneracy. (c) and (d) Two $\mathcal{T}$-broken topological phases in the presence of light irradiation. As the light intensity exceeds a critical value $k_{A}^{c}$, the system undergoes a topological phase transition from (c) the $\mathcal{T}$-broken TAI phase to (d) the QAH phase.}
    \label{fig1}
\end{figure}

Here, we propose a different strategy to design the QAH state without magnetic doping, thereby addressing the challenge for the realization of high-temperature QAH effects in realistic systems. The idea is to explore systems in which bulk topology is driven by the non-magnetic disorder and the spin degeneracy is lifted under irradiation of light fields.
Based on the Born approximation and Floquet theory, we demonstrate that the TAI exhibits rich spin-polarized topological phases with the $\mathcal{T}$-symmetry breaking and undergoes a sequence topological phase transitions from the TAI to $\mathcal{T}$-symmetry broken TAI, and then the QAH phase (see Fig. \ref{fig1}). In the absence of CPL, two spin sectors of the TAI host the same topological energy gap, exhibiting two spin edge channels with the Kramers degeneracy [see Fig. \ref{fig1}(b)]. When the incident CPL breaks the $\mathcal{T}$-symmetry, two spin sectors are no longer correlated and exhibit different responses. With increasing the light intensity, the spin splitting leads to two spin sectors possess different nontrivial energy gaps, resulting in the $\mathcal{T}$-symmetry broken TAI phase, in which one spin species edge state tends to penetrate into the bulk region due to its energy gap decreasing while the other spin species edge state is preserved [see Fig. \ref{fig1}(c)]. 
As the light intensity increases beyond a critical value $k_{A}^{c}$, the energy gap of one spin sector first closes and then reopens, while the other spin sector still possesses the nontrivial topology. In this case, the system evolves into the QAH phase with one gapless chiral edge channel [see Fig. \ref{fig1}(d)]. To distinguish these different topological phases, we compute the spin and charge Hall conductivities as functions light intensity and incident angle, which can be measured in experiments directly. At last, a comprehensive phase diagram characterized by the spin-dependent Chern numbers in a broad parameter regime is present.

To reveal the light-modulated topological phases mentioned above, we begin with the four-band Bernevig-Hughes-Zhang effective Hamiltonian for a 2D clean system with general type-II band alignment (e.g., HgTe/CdTe, InAs/GaSb quantum wells, or mori\'{e} superlattices)\cite{SCZhang,CXLiu,MsupL},
\begin{equation}
\begin{aligned}
&H_0(\mathbf{k})= \begin{pmatrix} h_{+}(\mathbf{k}) & 0 \\
0 & h_{-}(\mathbf{k}) \end{pmatrix},
\end{aligned}
\end{equation}
where $h_{s}(\mathbf{k}) = d_0(k)\tau_0+\mathbf{d}_{s}(\mathbf{k}) \cdot \boldsymbol{\sigma}$, and the index $s=\pm$ denotes spin $(\uparrow,\downarrow)$, or the inequivalent valleys $\pm K$ due to spin-valley locking in mori\'{e} superlattices. In the following, we take $s=\pm$ for spin-index. The lower diagonal block $h_{-}(\mathbf{k})$ is related to the upper diagonal block by the $\mathcal{T}$-symmetry $h_{-}(\mathbf{k})=h^{*}_{+}(-\mathbf{k})$.  The $\boldsymbol{\sigma}=(\sigma_x,\sigma_y,\sigma_z)$ are Pauli matrixes representing the pseudo-spin space. For the low-energy effective model to the lowest order in $k$, we have $d_0(k) = -Dk^2$ and $\mathbf{d}_{s}(\mathbf{k}) = (sAk_x,Ak_y,M-Bk^2)$, where the momentum vector is $\mathbf{k}=(k_x,k_y)$ and $k=\sqrt{k^2_x+k^2_y}$. $M$ is the Dirac mass term depicting the band inversion, the other parameters $A, B, D$ can be obtained from experiments \cite{QSHjpsj} and scaled by the lattice constant $a_L$, and we adopt a typical energy scale $\epsilon_0 \approx 73meV$ such that $A=a_L=1$, $B=-0.25$, $D = -0.2$. The spin Chern number can be expressed as $C_s = -\frac{s}{2}(\mathrm{sgn}(M)+\mathrm{sgn}(B))$, so the sign change of Dirac mass term indicates a topological phase transition from normal insulator (NI) ($M>0$) to TI ($M<0$).

Then, to achieve the TAI, we fix the Dirac mass to be $M>0$ and choose the uniformly distributed random potential modeled by $V(\boldsymbol{r}) = \sum_i U(\boldsymbol{R}_i)\delta(\boldsymbol{r}-\boldsymbol{R}_i)$,
where $U(\boldsymbol{R}_i)$ denotes the potential of impurity at position $\boldsymbol{R}_i$. In the momentum space, the impurity potential can be written as,
\begin{equation}
\begin{aligned}
V_{\boldsymbol{k'}\boldsymbol{k}}=\frac{1}{N}\sum_i U(\boldsymbol{R}_i) e^{-i(\boldsymbol{k}-\boldsymbol{k'})\cdot\boldsymbol{R}_i},
\end{aligned}
\end{equation}
which couples different momenta $\boldsymbol{k'}$ and $\boldsymbol{k}$ due to the translational symmetry breaking.

Let us consider the system under irradiation of CPL with the incident direction $\boldsymbol{n}=(\sin\theta\cos\varphi,\sin\theta\sin\varphi,\cos\theta)$ as illustrated in Fig. \ref{fig1}(a), where $\theta$ and $\varphi$ are the polar and azimuthal angles in the spherical coordinate. Using the Peierls substitution, the time-dependent Hamiltonian is given by $H_0(\mathbf{k})\rightarrow H_0(\mathbf{k}+e\mathbf{A(t)})$, in which the vector potential $\mathbf{A(t)}$ with amplitude $A_0$ and frequency $\Omega$ is expressed as $\mathbf{A(t)}=A_0[\cos(\Omega t)\mathbf{e_1}-\eta\sin(\Omega t)\mathbf{e_2}]$, and $\eta=+(-)$ correspond to the right- (left-) handed CPL. The two unit vectors $\mathbf{e_1}=(\cos\theta\cos\varphi,\cos\theta\sin\varphi,-\sin\theta)$ and $\mathbf{e_2}=(\sin\varphi,-\cos\varphi,0)$ satisfy orthogonal relations, i.e., $\boldsymbol{n}\cdot \mathbf{e_{1,2}}=0$ and $\mathbf{e_{1}}\cdot \mathbf{e_{2}}=0$. The time-periodic Hamiltonian can be expanded as $h_{s}(t,\mathbf{k})=\sum_{m} T_m(\mathbf{k})e^{im\omega t}$, and the expressions of $T_m(\mathbf{k})$ are concluded in Supplemental Material (SM) \cite{SM}. Based on the Floquet theory, we can derive a effective time-independent Hamiltonian in the off-resonant regime using the Magnus expansion \cite{Goldman},
\begin{equation}
\begin{aligned}
h^{\mathrm{eff}}_{s}(\mathbf{k}) &= T_0(\mathbf{k})+\sum_{m\geq 1}\frac{[T_m,T_{-m}]}{m\omega}+O(\frac{1}{\omega^2})\\
&= \tilde{d}_0(k)\tau_0+\tilde{\mathbf{d}}_{s}(\mathbf{k}) \cdot \boldsymbol{\sigma},
\end{aligned}
\label{eq:heff}
\end{equation}
where $\tilde{d}_0(k) = C-Dk^2$, and $C=-k^2_AD/2(1+\cos^2\theta)$ with light intensity $k_A = eA_0$. The light-modified
$\tilde{\mathbf{d}}_{s}(\mathbf{k}) = (sv_sk_x,v_sk_y,M_s-Bk^2)$, where $v_s=A-sA'$ with $A'=\eta 2ABk^2_A\cos\theta/\omega$. It is worth noting that light-renormalized Dirac mass
\begin{equation}
M_s=M-\Delta_s-k^2_AB/2(1+\sin^2\theta)
\label{eq:DM}
\end{equation}
with $\Delta_s=s\eta k^2_A A^2\cos\theta/\omega$ is spin-dependent, leading to two spin sectors with different responses to CPL. Besides, the azimuthal angle $\varphi$ is irrelevant to our problems.

The charge and spin Hall conductivities are important indicators of the topological phases.
For a disorder-free system, the Hall conductivity can be calculated by integrating the  Berry curvature in the momentum space. The presence of disorder breaks the translational symmetry and thus it is difficult to define Berry curvature since $\boldsymbol{k}$ is no longer a good quantum number. The alternative route to obtain the Hall conductivity is the Kubo-Bastin formula \cite{Bastin},
\begin{equation}
\begin{aligned}
&\sigma_{\mu\nu}(E_F,T) = \frac{ie^2\hbar}{\Omega}\int d\epsilon f(\epsilon)\times \\
&\mathrm{Tr}[(v_{\mu}\frac{dG^R(\epsilon)}{d\epsilon}v_{\nu}-v_{\nu}\frac{dG^R(\epsilon)}{d\epsilon}v_{\mu})
\delta(\epsilon-H)],
\end{aligned}
\label{eq:kb}
\end{equation}
where $\Omega$ is the volume of the system,  $f(\epsilon)=1/(e^{\epsilon-{E_F}/{k_BT}}+1)$ is the Fermi Dirac distribution with the Fermi level $E_F$ and temperature $T$, $H$ is the non-interacting Hamiltonian with velocity operators $v_{\mu} = \hbar^{-1}\partial_{k_{\mu}}H$, and the delta function is defined as $\delta(\epsilon-H) = -\frac{G^R(\epsilon)-G^A(\epsilon)}{2\pi i}$ where $G^{R(A)}$ is the retarded (advanced) Green¡¯s function.

Based on the effective Hamiltonian Eq.~(\ref{eq:heff}), the retarded Green's function of disordered systems can be defined as $G^R_V(\epsilon) = (\epsilon - h^{\mathrm{eff}}_{s} -V + i0^{+})$. After performing the disorder averaging, we have
\begin{equation}
\begin{aligned}
<G^R_V>_{dis}=G^R_s(\epsilon,\mathbf{k}) &= \frac{1}{\epsilon - h^{\mathrm{eff}}_{s}(\mathbf{k})  -\Sigma_{s}(\epsilon)},
\end{aligned}
\label{eq:gr}
\end{equation}
where the disorder induced self-energy $\Sigma_{s}(\epsilon)$ can be calculated by the self-consistent Born approximation (SCBA) \cite{SCBA-dos,Klier},
\begin{equation}
\begin{aligned}
\Sigma^R_{s}(\epsilon) &= \gamma\int\frac{d^2\mathbf{k}}{(2\pi)^2}\frac{1}{\epsilon - h^{\mathrm{eff}}_{s}(\mathbf{k}) -\Sigma^R_{s}(\epsilon)},
\end{aligned}
\label{eq:selfE}
\end{equation}
where $\gamma$ denotes the disorder strength (see the SM \cite{SM}). In order to calculate the conductivity using the Kubo-Bastin formula, we need to obtain the derivative of Green's function $\frac{dG(\epsilon)}{d\epsilon} = -G\frac{dG^{-1}(\epsilon)}{d\epsilon}G$,
where $\frac{dG^{-1}(\epsilon)}{d\epsilon} = 1-\frac{d\Sigma(\epsilon)}{d\epsilon}$.
The derivative of self-energy $\frac{d\Sigma(\epsilon)}{d\epsilon}$ can be obtained by differentiating the Eq.~(\ref{eq:selfE}),
\begin{equation}
\begin{aligned}
\frac{d\Sigma^R_{s}(\epsilon)}{d\epsilon} &= -\gamma\int\frac{d^2\mathbf{k}}{(2\pi)^2}G^R_s(\epsilon,\mathbf{k})(1-\frac{d\Sigma^R_{s}(\epsilon)}{d\epsilon})G^R_s(\epsilon,\mathbf{k})
\end{aligned}
\label{eq:dselfE}
\end{equation}
Then, we expand the self-energy into Pauli matrices $\Sigma = \Sigma_0\sigma_0+\Sigma_z\sigma_z$, where the coefficients $\Sigma_{i}$ and $d\Sigma_{i}/d\epsilon$ are determined self-consistently by combining Eqs.~(\ref{eq:gr}), (\ref{eq:selfE}) and (\ref{eq:dselfE}). We present the details of calculation in SM \cite{SM}. After obtaining the Green's function $G^{R(A)}(\epsilon)$ and $dG^{R(A)}(\epsilon)/d\epsilon$, we can calculate the longitudinal and transverse (Hall) conductivity using the Kubo formalism Eq.~(\ref{eq:selfE}).

Though the above approach provides a possible pathway to calculate transport properties of disorder systems, the delta function $\delta(\epsilon-H)$ in Eq.~(\ref{eq:kb}) causes difficulty in the numerical integration \cite{PhysRevB.64.014416}. The traditional way to separate anomalous Hall transports from the Kubo-Bastin formula is given by Streda-Smrcka \cite{Streda-Smrcka1,Streda-Smrcka2}. Unfortunately, the Streda-Smrcka decomposition contains an overlap, which makes it difficult to calculate the intrinsic Hall conductivity \cite{Kontani}. More recently, Bonbien and Manchon \cite{SD-KB} proposed a new decomposition $\sigma_{\mu\nu} = \sigma^{surf}_{\mu\nu}+\sigma^{sea}_{\mu\nu}$, in which the Fermi sea term $\sigma^{sea}_{\mu\nu}$ and Fermi surface term $\sigma^{surf}_{\mu\nu}$ are expressed in terms of Green's functions as
\begin{equation}
\begin{aligned}
\sigma^{surf}_{\mu\nu} &= \frac{e^2\hbar}{4\pi\Omega}\int d\epsilon \frac{df(\epsilon)}{d\epsilon}\times\\
& \mathrm{Re}\{\mathrm{Tr}[v_{\mu}(G^R(\epsilon)-G^A(\epsilon))v_{\nu}(G^R(\epsilon)-G^A(\epsilon))]\},\\
\sigma^{sea}_{\mu\nu} &= \frac{e^2\hbar}{2\pi\Omega}\int d\epsilon f(\epsilon)\times\\
&\mathrm{Re}\{\mathrm{Tr}[v_{\mu}(G^R(\epsilon)-G^A(\epsilon))v_{\nu}(\frac{dG^R(\epsilon)}{d\epsilon}+\frac{dG^A(\epsilon)}{d\epsilon})]\}.
\end{aligned}
\label{eq:newd}
\end{equation}

\begin{figure}[htb]
\setlength{\belowcaptionskip}{-0.20cm}
 \setlength{\abovecaptionskip}{-0.10cm}
\includegraphics[scale=0.43]{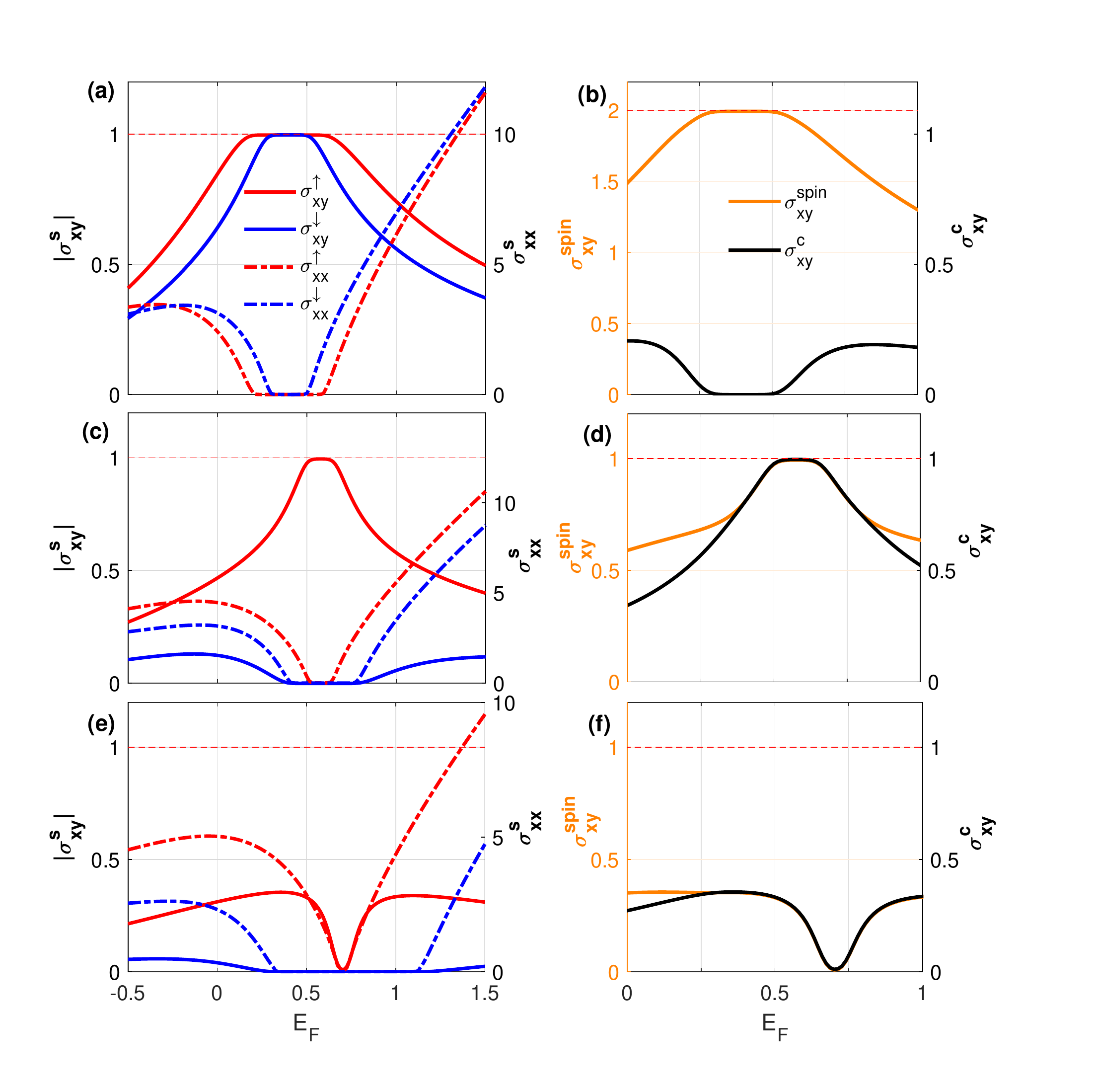}
\caption{The conductivities of TAI with disorder strength $\gamma = 0.1$ under irradiation of CPL versus the Fermi level $E_F$. The light intensity is (a) and (b) $k_A = 0.75$, (c) and (d) $k_A = 1.25$,  (e) and (f) $k_A = 1.6$. In the left panels, the spin-dependent Hall conductivities $\sigma^{s}_{xy}$ (solid line) and longitudinal conductivity $\sigma^{s}_{xx}$ ( dashed line), where $s=[\uparrow(\rm{red}),\downarrow(\rm{blue})]$. In the right panels, the spin Hall conductivity $\sigma^{spin}_{xy}$  (orange solid line)
and the charge Hall conductivity $\sigma^{c}_{xy}$ (black solid line). The charge conductivity is in the unit of $\frac{e^2}{h}$ and the spin conductivity is in the unit of $\frac{e}{4\pi}$.}
\label{Fig:FTAI}
\end{figure}

The advantage of this new decomposition in Eq.~(\ref{eq:newd}) is that the Fermi surface term $\sigma^{surf}_{\mu\nu}$ is directly recognized as the Kubo-Greenwood formula, and the Fermi sea term $\sigma^{surf}_{\mu\nu}$ is related to intrinsic Hall conductivity contributed by the Berry curvature \cite{SD-KB}.
As expected, in our calculation,
the Hall conductivity $\sigma^{sea}_{xy}$ only accepts contributions from the Fermi sea term and the longitudinal conductivity $\sigma^{surf}_{xx}$ is solely determined by the Fermi surface term. Therefore, we can ignore the upper decomposing index in the following.
As shown in Fig. S2 in the SM \cite{SM}, the new decomposition of Kubo-Bastin formula can well reproduce topological features of the TAI obtained from the Landauer-B\"{u}ttiker formalism \cite{TAI}, providing a reliable and effective approach to investigate the disordered topological phases.

Next, we turn to study the TAI subject to CPL. The strength of disorder is fixed at $\gamma = 0.1$, which makes the system initially to be the TAI.  It is worth noting that the Dirac mass $M_s$ [see Eq.~(\ref{eq:DM})] without disorder always monotonously increases under the irradiation of CPL [see Fig. S1 in the SM \cite{SM}], i.e., the CPL always gives positive correction to $M_s$. Therefore, there is no topological phase transition if the system is initially a normal insulator (NI) ($M_s>0$). The importance of non-magnetic disorder is to ensure that the initial state is the TAI ($M_s<0$). The topological phase transition occurs when the Dirac mass is individually from $M_s<0$ to $M_s>0$.
Once irradiation of CPL is applied, two spin sectors exhibit different responses since the spin-resolved coupling destroys the $\mathcal{T}$-symmetry. To identify the spin-polarized topological phases with spin-resolved Chern fluxes ($C^{\uparrow}$, $C^{\downarrow}$), it is required to calculate the spin-dependent conductivities $\sigma^{s}_{xx}$ and $\sigma^{s}_{xy}$ with $s=(\uparrow,\downarrow)$. In addition, the global topological features can also be characterized by the spin Hall conductivity $\sigma^{spin}_{xy}$ and charge Hall conductivity $\sigma^{c}_{xy}$, which are respectively defined as \cite{condSC},
\begin{equation}
\begin{aligned}
\sigma^{spin}_{xy} &= \frac{\hbar}{2e}(\sigma^{\uparrow}_{xy}-\sigma^{\downarrow}_{xy}),\\
\sigma^{c}_{xy} &= \sigma^{\uparrow}_{xy}+\sigma^{\downarrow}_{xy}.
\end{aligned}
\label{eq:condsc}
\end{equation}

We first consider the cases of normal incidence (i.e., $\theta = 0$) and right-handed CPL (i.e., $\eta = 1$).
In the left panels of Fig. \ref{Fig:FTAI}, we plot the spin-dependent conductivity $\sigma^{s}_{xx}$ and $\sigma^{s}_{xy}$ as a function of the Fermi level. For convenience, we employ  $\sigma^{\uparrow}_{xy}=|\sigma^{\uparrow}_{xy}|$ and $\sigma^{\downarrow}_{xy}=-|\sigma^{\downarrow}_{xy}|$. For a weak light intensity ($k_A = 0.75$) of CPL, we find that the light-induced spin-splitting gives rise to different nontrivial gaps of two spin sectors. At a given Fermi level, the Hall conductivities $\sigma^{\uparrow}_{xy}$ and $\sigma^{\downarrow}_{xy}$ are quantized to $e^2/h$ and -$e^2/h$  in the energy gap [see Fig. \ref{Fig:FTAI}(a)], respectively, corresponding to $C_{\uparrow}=1$ and $C_{\downarrow}=-1$. In this case, the system only contributes to a net transport of spin ($\sigma^{spin}_{xy} = 2\frac{e}{4\pi}$) but no net charge transfer ($\sigma^{c}_{xy} = 0$) [see Fig. \ref{Fig:FTAI}(b)], resulting in the $\mathcal{T}$-broken TAI phase. With increasing the light intensity $k_A$, the evolution of spin-down sector is more rapid than that of spin-up sector. Until $k_A$ is beyond a critical value $k_A^c$, at which the energy gap of spin-down sector closes and reopens, the spin-up sector still possesses the non-trivial (or inverted) bulk topology. As shown in Fig. \ref{Fig:FTAI}(c), we can see that the Hall conductivities $\sigma^{\uparrow}_{xy}= e^2/h$ and $\sigma^{\downarrow}_{xy}=0$ at the light-intensity $k_A = 1.25$, indicating that the spin-down sector is topologically trivial ($C_{\downarrow}=0$) and the spin-up sector is topologically nontrivial ($C_{\uparrow}=1$). In this case, a QAH phase with charge Hall conductivity $\sigma^{c}_{xy}=e^2/h$ is present [see Fig. \ref{Fig:FTAI}(d)]. The further increase of $k_A$ will close the gap of spin-up sector, leading to a transition from the QAH phase into NI, as shown in Figs. \ref{Fig:FTAI}(e) and \ref{Fig:FTAI}(f).

\begin{figure}[htb]
 \setlength{\belowcaptionskip}{-0.2cm}
 \setlength{\abovecaptionskip}{-0.0cm}
\includegraphics[scale=0.55]{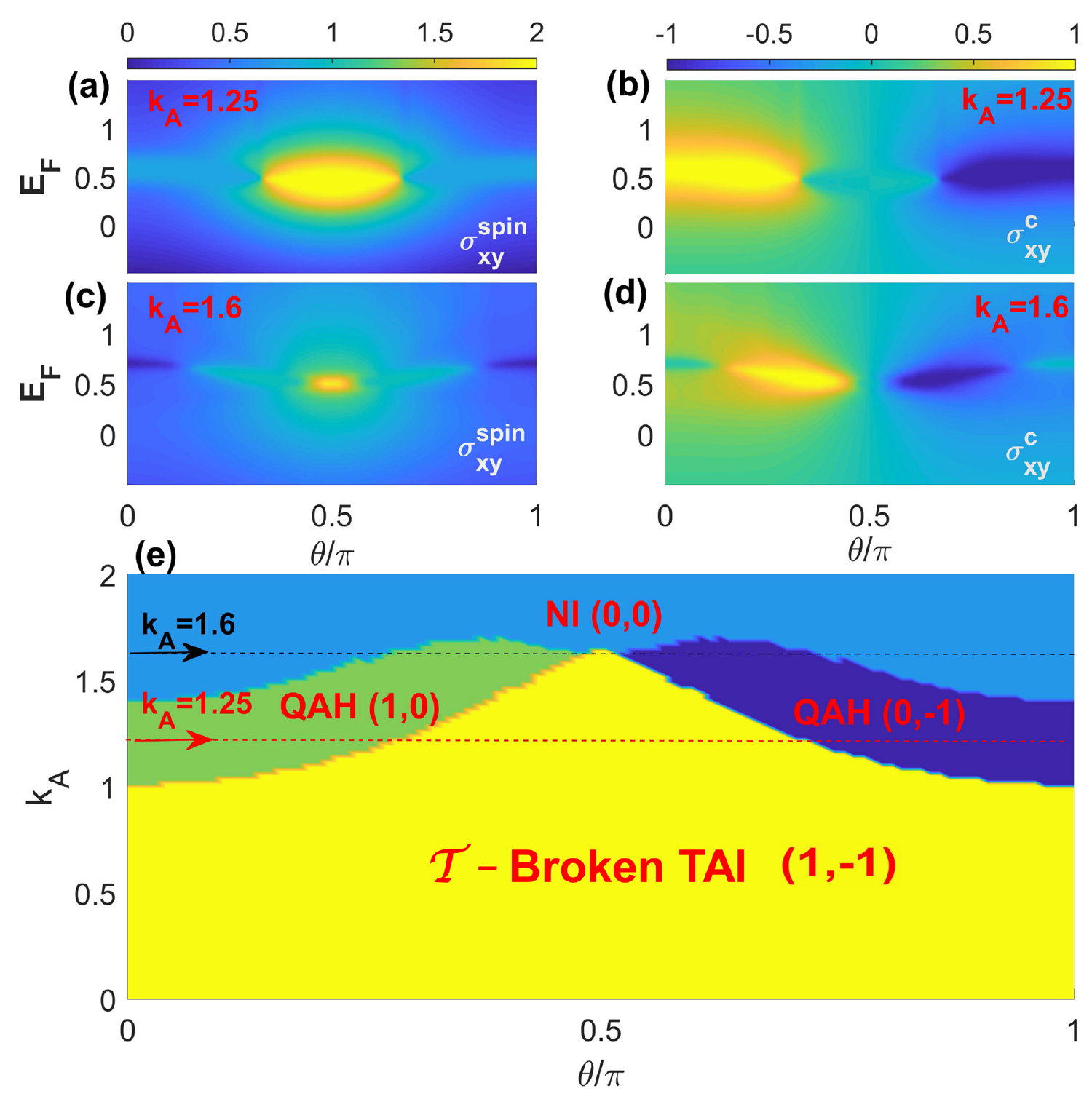}
\caption{(a) and (c) The spin Hall conductivity, and (b) and (d) charge Hall conductivity of TAI with disorder strength $\gamma = 0.1$ as a function of the Fermi level $E_F$ and polar angle $\theta$. The color map represents the value of $\sigma^{spin/c}_{xy}$ obtained from Eq.(\ref{eq:newd}). (e) The phase diagram of the TAI under CPL on the $k_A-\theta$ parameter plane depicted by the spin-dependent Chern fluxes $(C^{\uparrow},C^{\downarrow})$. The horizontal red and black dashed lines in the panel (e) denote the light intensity of CPL $k_A = 1.25$ and $k_A = 1.6$, respectively.}
\label{Fig:pdiagram}
\end{figure}

Beyond the topological phase transitions driven by the light-intensity $k_A$, it is worth noting that the light-renormalized Dirac mass $M_s$ in Eq.~(\ref{eq:DM}) is dependent on the polar angle $\theta$ of incident CPL. Therefore, a further study on topological features of TAI as a function of $\theta$ is also desirable. As shown in Figs. \ref{Fig:pdiagram}(a)-\ref{Fig:pdiagram}(d), the change of polar angle $\theta$ at the fixed $k_A$ realizes the transition between $\mathcal{T}$-broken TAI phase and QAH phase. We also find that the QAH phase in the region of $\theta < \frac{\pi}{2}$ (right-handed CPL) and that in the region of $\theta > \frac{\pi}{2}$ (left-handed CPL) exhibit the opposite value of charge Hall conductivity $\sigma^{c}_{xy}$, which respectively corresponds to the spin-resolved Chern fluxes ($C^{\uparrow}=1$, $C^{\downarrow}=0$) and ($C^{\uparrow}=0$, $C^{\downarrow}=-1$). This implies that the polarization of CPL can control the spin polarization and propagating direction of edge states. To fully understand light-induced spin-polarized topological phases, we present a comprehensive study on spin-resolved Chern fluxes ($C^{\uparrow}$, $C^{\downarrow}$) in a broad parameter regime. 
As shown in Fig. \ref{Fig:pdiagram}(e), the phase diagram on the $\theta - k_A$ parameter plane depicts that there are three distinct spin-polarized phases, such as the $\mathcal{T}$-broken TAI phase with ($C^{\uparrow}=1$, $C^{\downarrow}=-1$), the QAH phase with ($C^{\uparrow}=1$, $C^{\downarrow}=0$) or ($C^{\uparrow}=0$, $C^{\downarrow}=-1$), and the NI phase with ($C^{\uparrow}=0$, $C^{\downarrow}=0$). In addition, it is worth noting that the spin-splitting term $\Delta=0$ in Eq.(\ref{eq:DM}) at $\theta=\frac{\pi}{2}$ leads to light-renormalized Dirac mass $M_{\uparrow} = M_{\downarrow}$ and thus the QAH phase vanishes.

In summary, using the effective Hamiltonian based on the Floquet theory and Born approximation, we have theoretically investigated the spin-polarized topological phases and inevitable QAH state under the interplay of light and non-magnetic disorder. The $\mathcal{T}$-symmetry broken TAI phase and QAH phase can be tuned via light intensity and polar angle. Using the Green's function method and the new decomposition of Kubo-Bastin formula, the topological phase transition coupled with the light intensity and the incident angle has been systematically studied, and a comprehensive phase diagram characterized by the spin-dependent Chern fluxes ($C^{\uparrow}$, $C^{\downarrow}$) is present. To date, this is the only phase diagram found to depict the spin-polarized topological properties in a non-magnetic disorder system subject to CPL. In addition, this study can also easily extend to valley-polarized topological states, such as mori\'{e} superlattices. Therefore, our work provides a reliable avenue to study light-driven topological phases with $\mathcal{T}$-symmetry breaking in widely non-magnetic disorder systems. Furthermore, the QAH state induced by CPL, without magnetic doping, can survive at high temperatures, and thus are important for applications of topological spintronics devices.

\begin{acknowledgments}
This work was supported by the National Natural Science Foundation of China (NSFC, Grants No. 11974062, No. 12047564, No. 11704177, No. 12074108 and No. 11704106), the Chongqing Natural Science Foundation (Grants No. cstc2019jcyj-msxmX0563), the Fundamental Research Funds for the Central Universities of China (Grant No. 2020CDJQY-A057), and the Beijing National Laboratory for Condensed Matter Physics.
\end{acknowledgments}


%

\newpage

\begin{widetext}
\newpage

\setcounter{figure}{0}
\setcounter{equation}{0}
\makeatletter

\makeatother
\renewcommand{\thefigure}{S\arabic{figure}}
\renewcommand{\thetable}{S\Roman{table}}
\renewcommand{\theequation}{S\arabic{equation}}

\begin{center}
	\textbf{
		\large{Supplemental Material for}}
	\vspace{0.2cm}
	
	\textbf{
		\large{
			``Photoinduced Quantum Anomalous Hall States in the Topological Anderson Insulator" }
	}
\end{center}

In this Supplemental Material, we present the detailed derivation of the Floquet Hamiltonian and Green's function using self-consistent Born approximation.

\section{The Floquet Hamiltonian}
We start from four-band Bernevig-Hughes-Zhang effective Hamiltonian for a 2D clean system with general type-II band alignment as
\begin{equation}
\begin{aligned}
h_{s}(\mathbf{k}) = d_0(k)\sigma_0+\mathbf{d}_{s}(\mathbf{k}) \cdot \boldsymbol{\sigma},
\end{aligned}
\end{equation}
where $h_{+}$ and $h_{-}$ respectively represent the upper and lower block with $s=\pm$ for $(\uparrow,\downarrow)$ as described in Eq.(1) in the main text. In the presence of circularly polarized light (CPL), the time-dependent Hamiltonian is given by
$h_{s}(\mathbf{k})\rightarrow h_{s}[\mathbf{k}+e\mathbf{A(t)}]$, which can be expanded as
$h_{s}(t,\mathbf{k})=\sum_{m} T_m(\mathbf{k})e^{im\omega t}$. Here, the Fourier components $T_m$ are

\begin{equation}
\begin{aligned}
T_0(\mathbf{k})  &= (-k^2_AD+k^2_AD/2\sin^2\theta-Dk^2)\sigma_0
+(M-k^2_AB+k^2_AB/2\sin^2\theta-Bk^2)\sigma_z
+sAk_x\sigma_x+Ak_y\sigma_y,\\
\\
T_{\pm 1}(\mathbf{k}) &= [Dk_A(\mp ik_y\eta-k_x\cos\theta)\cos\varphi
+Dk_A(\pm ik_x\eta-k_y\cos\theta)\sin\varphi]\sigma_0\\
&+sA/2k_A(\cos\theta\cos\varphi\mp i\eta\sin\varphi)\sigma_x
+ A/2k_A(\cos\theta\sin\varphi\pm i\eta\cos\varphi)\sigma_y\\
&+[Bk_A(\mp ik_y\eta-k_x\cos\theta)\cos\varphi
+Bk_A(\pm ik_x\eta-k_y\cos\theta)\sin\varphi]\sigma_z,\\
\\
T_{\pm 2}(\mathbf{k}) &= -D/8k^2_A(-1+\cos\theta)\sigma_0
-B/8k^2_A(-1+\cos\theta)\sigma_z,
\end{aligned}
\label{seq:Tm}
\end{equation}
and $T_m = 0$ for $m\ge2$. In the off-resonant approximation, we can derive a effective time-independent Hamiltonian using the Magnus expansion~\cite{ZBYan}
\begin{equation}
\begin{aligned}
h^{\mathrm{eff}}_{s}(\mathbf{k}) &= T_0(\mathbf{k})+\sum_{m\geq 1}\frac{[T_m,T_{-m}]}{m\omega}+O(\frac{1}{\omega^2}).
\end{aligned}
\label{seffSM}
\end{equation}
The commutation relation in Eq.(\ref{seffSM}) gives
\begin{equation}
\begin{aligned}
\mathrm{[}T_1,T_{-1}] & = -2ABk_A^2k_x\eta\cos\theta\tau_x - 2sABk_A^2k_y\eta\cos\theta\tau_y - sA^2k_A^2\eta\cos\theta\tau_z,\\
\mathrm{[}T_2,T_{-2}] & = 0,
\end{aligned}
\end{equation}
and then we can obtain the effective Floquet Hamiltonian as
\begin{equation}
\begin{aligned}
h^{\mathrm{eff}}_{s}(\mathbf{k}) = \tilde{d}_0(k)\sigma_0+\tilde{\mathbf{d}}_{s}(\mathbf{k}) \cdot \boldsymbol{\sigma},
\end{aligned}
\label{seq:heff}
\end{equation}
where $\tilde{d}_0(k) = C-Dk^2$, and $C=-k^2_AD/2(1+\cos^2\theta)$ with light intensity $k_A = eA_0$. The renormalized
$\tilde{\mathbf{d}}_{s}(\mathbf{k})$ is expressed as  $\tilde{\mathbf{d}}_{s}(\mathbf{k})= (sv_sk_x,v_sk_y,M_s-Bk^2)$, where $v_s=A-sA'$ with $A'=\eta 2ABk^2_A\cos\theta/\omega$. The renormalized Dirac mass $M_s$ is spin-dependent and expressed as
\begin{equation}
\begin{aligned}
M_s &= M + f_s(\theta) k^2_A,\\
f_s(\theta) &= -sA^2\cos\theta/\omega-B/2(1+\sin^2\theta).
\end{aligned}
\label{seq:DM}
\end{equation}
Here, the first term in $f_s(\theta)$ is very small compared to the second term in the high frequency regime. In this case, the value of the function $f_s(\theta)$ is always positive $(B<0)$ and the CPL gives positive correction to $M_s$ [see Fig. \ref{Fig:DM}]. It is worth noting that $M_s$ [see Eq.(\ref{seq:DM})] always monotonously increases with the intensity of CPL ($k_A$). Therefore, there is no topological phase transition if the system is initially a normal insulator (NI) ($M>0$).
\
\begin{figure}[htb]
\setlength{\belowcaptionskip}{0cm}
 \setlength{\abovecaptionskip}{0cm}
\includegraphics[scale=0.38]{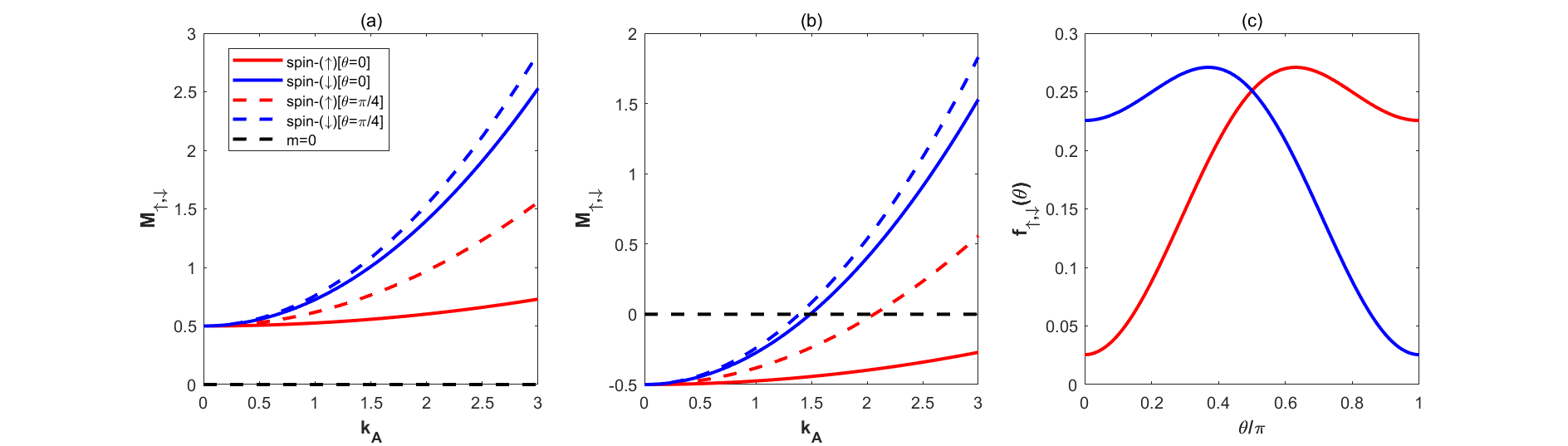}
\caption{The renormalized Dirac mass $M_s$ as function of light intensity $k_A$ for (a) $M = 0.5$ and (b) $M = -0.5$. (b) The angle-dependent function $f_s(\theta)$.  }
\label{Fig:DM}
\end{figure}

\section{The Green's function}
In this section, we calculate the Green's function based on the effective Floquet Hamiltonian $h^{\mathrm{eff}}_{s}(\mathbf{k})$ [see Eq.(\ref{seq:heff})] after on-site magnetic disorder is considered. The retarded Green's function with disorder can be defined as $G^R_V(\epsilon) = (\epsilon - h^{\mathrm{eff}}_{s} -V + i0^{+})$. The random potential modeled by $V(\boldsymbol{r}) = \sum_i U(\boldsymbol{R}_i)\delta(\boldsymbol{r}-\boldsymbol{R}_i)$, where $U(\boldsymbol{R}_i)$ denotes the potential of impurity at position $\boldsymbol{R}_i$, which uniformly distributed in the interval $[-W/2,W/2]$. The disorder averaged Green's function $<G^R_V>_{dis}= G^R$ [i.e., Eq.(5) of the main text] is determined by the self energy $\Sigma_{s}(\epsilon)$, which can be calculated by the SCBA \cite{SCBA-dos},
\begin{equation}
\begin{aligned}
\Sigma^R_{s}(\epsilon) &= \gamma\int\frac{d^2\mathbf{k}}{(2\pi)^2}\frac{1}{\epsilon - h^{\mathrm{eff}}_{s}(\mathbf{k}) -\Sigma^R_{s}(\epsilon)},
\end{aligned}
\label{seq:selfE}
\end{equation}
where the strength of disorder is defined as $\gamma = \frac{W^2}{12}$. Then, we decompose the self energy into Pauli matrices $\Sigma^R_{s} = \Sigma^R_{s(0)}\sigma_0+\Sigma^R_{s(x)}\sigma_x+\Sigma^R_{s(y)}\sigma_y+\Sigma^R_{s(z)}\sigma_z$. From the first-order Born approximation, we know that the $\Sigma^R_{s(0)}$ and $\Sigma^R_{s(z)}$ give corrections to the energy $\epsilon$ and mass term $M_s$, respectively. The other two components are $\Sigma^R_{s(x,y)} = 0$ and can be ignored in the following self-consistent calculation. By performing the integration in Eq.(\ref{seq:selfE}), we can obtain the expression of $\Sigma^R_{s(0)}$ and $\Sigma^R_{s(z)}$ as
\begin{equation}
\begin{aligned}
\Sigma^R_{s(0)}(\epsilon) &= \gamma\int\frac{kdk}{(2\pi)}\frac{ Dk^2 + z_R}{(-B^2 + D^2)k^4 - m_R^2 + z_R^2 + k^2 (2Bm_R - v_s^2 + 2Dz_R)},\\
\Sigma^R_{s(z)}(\epsilon) &= \gamma\int\frac{kdk}{(2\pi)}\frac{-Bk^2 + m_R}{(-B^2 + D^2)k^4 - m_R^2 + z_R^2 + k^2 (2Bm_R - v_s^2 + 2Dz_R)},
\end{aligned}
\label{seq:selfEeq}
\end{equation}
where we define $z_R = \epsilon - \tilde{d}_0(k)-\Sigma^R_{s(0)}(\epsilon)$ and $m_R = M_s -\Sigma^R_{s(z)}(\epsilon)$.
Then, we turn to calculate the derivative of the Green's function $\frac{dG(\epsilon)}{d\epsilon}$. By differentiating the identity $G(\epsilon)G^{-1}(\epsilon) = 1$, we can establish a connection between $\frac{dG(\epsilon)}{d\epsilon}$ and $\frac{d\Sigma(\epsilon)}{d\epsilon}$
\begin{equation}
\begin{aligned}
\frac{dG(\epsilon)}{d\epsilon} &= -G(\epsilon)\frac{dG^{-1}(\epsilon)}{d\epsilon}G(\epsilon)\\
&= -G(\epsilon)(1-\frac{d\Sigma(\epsilon)}{d\epsilon})G(\epsilon).
\end{aligned}
\label{seq:selfEeq}
\end{equation}
The derivative of self energy $\frac{d\Sigma(\epsilon)}{d\epsilon}$ can be obtained by taking the derivative of Eq.(\ref{seq:selfE})
\begin{equation}
\begin{aligned}
 \frac{d\Sigma^R_{s}(\epsilon)}{d\epsilon}&= -\gamma\int\frac{d^2\mathbf{k}}{(2\pi)^2}G^R_s(\epsilon,\mathbf{k})(1-\frac{d\Sigma^R_{s}(\epsilon)}{d\epsilon})G^R_s(\epsilon,\mathbf{k}).
\end{aligned}
\label{seq:dselfE}
\end{equation}
After a straightforward calculation of the right hand side of Eq.(\ref{seq:dselfE}), the self-consistent equations of $\frac{d\Sigma^R_{s}(\epsilon)}{d\epsilon}$ can be expressed as
\begin{equation}
\begin{aligned}
 \frac{d\Sigma^R_{s(0)}(\epsilon)}{d\epsilon}&=
 -\gamma\int\frac{kdk}{(2\pi)}\frac{ a_4k^4 + a_2k^2 + a_0}{[(-B^2 + D^2)k^4 - m_R^2 + z_R^2 + k^2 (2Bm_R - v_s^2 + 2Dz_R)]^3},\\
  \frac{d\Sigma^R_{s(z)}(\epsilon)}{d\epsilon}&=
 -\gamma\int\frac{kdk}{(2\pi)}\frac{ b_4k^4 + b_2k^2 + b_0}{[(-B^2 + D^2)k^4 - m_R^2 + z_R^2 + k^2 (2Bm_R - v_s^2 + 2Dz_R)]^3},
\end{aligned}
\label{eq:dselfEeq}
\end{equation}
where the coefficient $a_{4,2,0}$ and $b_{4,2,0}$ are defined as
\begin{equation}
\begin{aligned}
a_4 &= (B^2+D^2)(1-\frac{d\Sigma^R_{s(0)}(\epsilon)}{d\epsilon})
+ 2BD\frac{d\Sigma^R_{s(z)}(\epsilon)}{d\epsilon},\\
a_2 &= v_s^2(1-\frac{d\Sigma^R_{s(0)}(\epsilon)}{d\epsilon})
+ 2D[-m_R\frac{d\Sigma^R_{s(z)}(\epsilon)}{d\epsilon} + z_R(1-\frac{d\Sigma^R_{s(0)}(\epsilon)}{d\epsilon})]
- 2B[m_R(1-\frac{d\Sigma^R_{s(0)}(\epsilon)}{d\epsilon}) - z_R\frac{d\Sigma^R_{s(z)}(\epsilon)}{d\epsilon})],\\
a_0 &=  -2m_Rz_R\frac{d\Sigma^R_{s(z)}(\epsilon)}{d\epsilon} + (m_R^2 + z_R^2)(1-\frac{d\Sigma^R_{s(0)}(\epsilon)}{d\epsilon}),\\
b_4 &= -(B^2+D^2)\frac{d\Sigma^R_{s(z)}(\epsilon)}{d\epsilon}
- 2BD(1-\frac{d\Sigma^R_{s(0)}(\epsilon)}{d\epsilon}),\\
b_2 &= v_s^2\frac{d\Sigma^R_{s(z)}(\epsilon)}{d\epsilon}
- 2B[-m_R\frac{d\Sigma^R_{s(z)}(\epsilon)}{d\epsilon} + z_R(1-\frac{d\Sigma^R_{s(0)}(\epsilon)}{d\epsilon})]
+ 2D[m_R(1-\frac{d\Sigma^R_{s(0)}(\epsilon)}{d\epsilon}) - z_R\frac{d\Sigma^R_{s(z)}(\epsilon)}{d\epsilon})],\\
b_0 &=  2m_Rz_R(1-\frac{d\Sigma^R_{s(0)}(\epsilon)}{d\epsilon})  - (m_R^2 + z_R^2)\frac{d\Sigma^R_{s(z)}(\epsilon)}{d\epsilon}.
\end{aligned}
\end{equation}

After obtaining the Green's function $G^{R(A)}(\epsilon)$ and $dG^{R(A)}(\epsilon)/d\epsilon$, we apply the new decomposition of Kubo-Bastin formula to reproduce the formation of TAI without light irradiation
\begin{equation}
\begin{aligned}
\sigma^{surf}_{\mu\nu} &= \frac{e^2\hbar}{4\pi\Omega}\int d\epsilon \frac{df(\epsilon)}{d\epsilon} \mathrm{Re}\{\mathrm{Tr}[v_{\mu}(G^R(\epsilon)-G^A(\epsilon))v_{\nu}(G^R(\epsilon)-G^A(\epsilon))]\},\\
\sigma^{sea}_{\mu\nu} &= \frac{e^2\hbar}{2\pi\Omega}\int d\epsilon f(\epsilon)
\mathrm{Re}\{\mathrm{Tr}[v_{\mu}(G^R(\epsilon)-G^A(\epsilon))v_{\nu}(\frac{dG^R(\epsilon)}{d\epsilon}+\frac{dG^A(\epsilon)}{d\epsilon})]\}.
\end{aligned}
\label{seq:newd}
\end{equation}

\begin{figure}[htb]
\setlength{\belowcaptionskip}{0cm}
 \setlength{\abovecaptionskip}{0cm}
\includegraphics[scale=0.7]{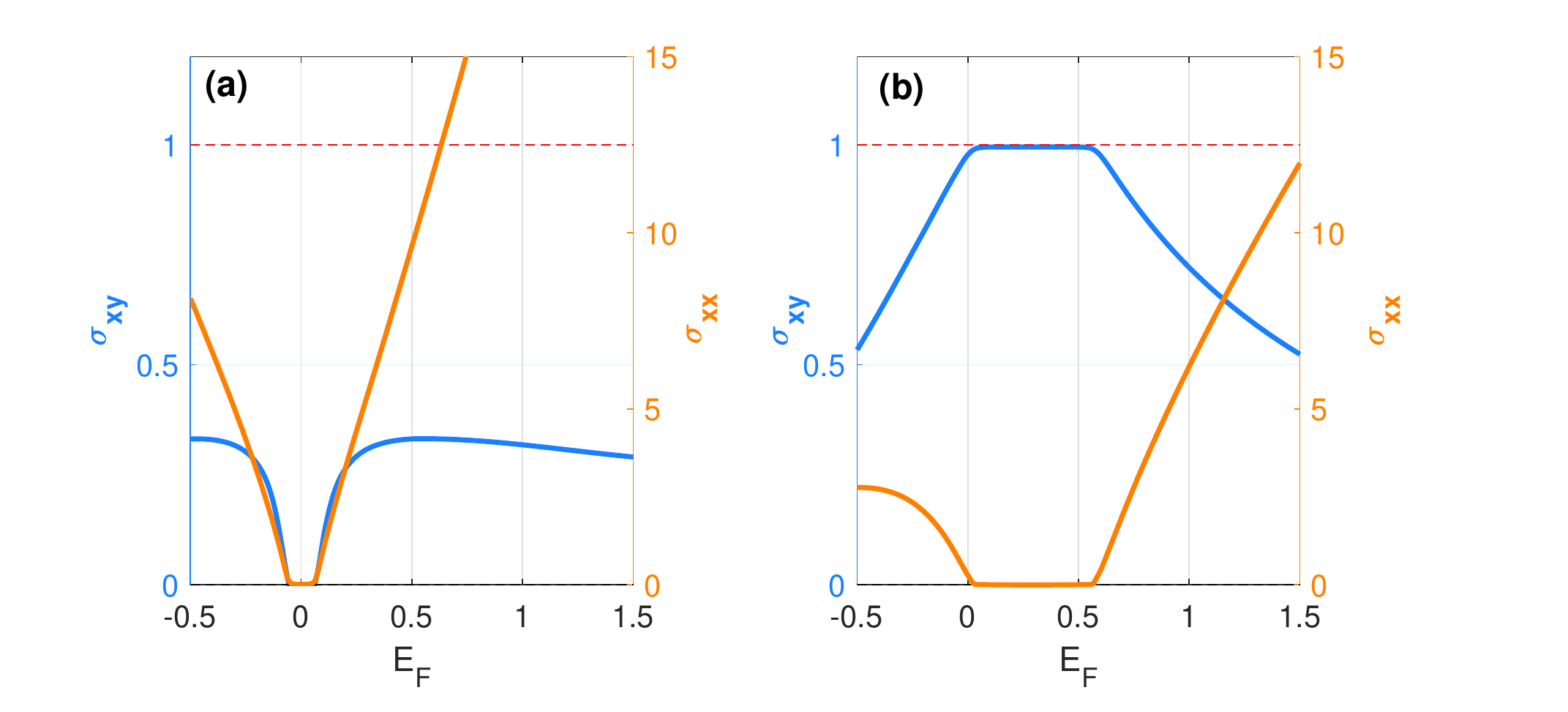}
\caption{ The Hall conductivity $\sigma_{xy}$ (blue) of the spin-up block $h_{\uparrow}(\mathbf{k})$  and longitudinal conductivity $\sigma_{xx}$ (orange) of disordered BHZ Hamiltonian in the absence of light irradiation as a function of the Fermi level $E_F$. The strength of disorder is at (a) $\gamma = 0.01$  and (b) $\gamma = 0.1$.
The conductivity is in the unit of $\frac{e^2}{h}$.}
\label{Fig:TAI}
\end{figure}

In the absence of light field (i.e., $k_A = 0$), the Kramers degeneracy indicates that the spin index can be discarded due to the presence of $\mathcal{T}$-symmetry. Without loss of generality, the initial positive value of the Dirac mass is set to $M=0.07$. For a very weak disorder potential $\gamma = 0.01$, the renormalized Dirac mass is still positive and the system is a normal insulator (NI), showing zero value of Hall conductivity of the spin-up block  inside the band gap [see Fig. \ref{Fig:TAI}(a)]. With increasing the disorder strength, the Dirac mass can be expectedly reduced and further the band inversion occurs. In this case, the Hall conductivity $\sigma_{xy}$ of the spin-up block is quantized to be $e^2/h$ in the band gap [see Fig. \ref{Fig:TAI}(b)]; that is, the system turns to the TAI. Therefore, the new decomposition of Kubo-Bastin formula can well reproduce the topological features obtained from the Landauer-B\"{u}ttiker formalism, providing a reliable and effective approach to investigate the disordered topological phases.

\end{widetext}

\end{document}